[1994/12/01]
\documentclass[english]{article}
\usepackage{textcomp}
\usepackage{amssymb,amsmath,amsthm}
\usepackage[all]{xy}

\usepackage{graphicx}
\usepackage[english]{babel}
\newcommand{\wrd}{\Rightarrow}

\newcommand{\M}{\mathfrak{M}}
\newcommand{\F}{\mathfrak{F}}
\newcommand{\C}{\mathfrak{C}}

\newcommand{\R}{\mathcal{R}}
\newcommand{\I}{\mathcal{I}}
\newcommand{\D}{\mathcal{D}}
\newcommand{\G}{\mathcal{G}}

\newcommand{\ro}{\varrho_1{\small ,\ldots,}\varrho_n}

\newcommand{\union}{\cup}

\newcommand{\cons}[1]{\text{{\tt '#1'}}}

\newcommand{\da}{\downarrow}

\theoremstyle{plain}
\newtheorem{thm}{Theorem}[section]
\newtheorem{lem}[thm]{Lemma}
\newtheorem{prop}[thm]{Proposition}
\theoremstyle{definition}
\newtheorem{defn}{Definition}[section]

\newtheorem{exmp}{Example}[section]
\theoremstyle{remark}

\title{Translating a first-order modal language\\to relational algebra}
\author{Yeb Havinga\thanks{Work supervised by Balder ten Cate, Institute for Logic, Language and Computation, Universiteit van Amsterdam.}\\Portavita B.V., The Netherlands}
\date{\today}

\providecommand{\etc}[1]{etc.}

\begin{document}
\maketitle

\section{Introduction}
\defn[Kripke structure]
A \emph{Kripke structure} is a tuple $\F$ whose first component is a non-empty set $\G$ 
called the \emph{universe} of $\F$ and whose remaining components are binary relations on $\G$. We assume that every
Kripke structure has at least one relation.
\\
\\
\noindent This paper is about Kripke structures that are 
\begin{enumerate}
\item inside a relational database.
\item queried with a \emph{modal} language.
\end{enumerate}
At first the modal language that is used is introduced,
followed by a definition of the database and relational algebra.
Based on these definitions two things are described:
\begin{enumerate}
\item a mapping from components of the model structure to a relational database schema and instance.
\item a translation from queries in the modal language to relational algebra queries.
\end{enumerate}

\section{The modal language}

\subsection{Language}
The modal language used is an adaptation of the language used in \cite{fit_book}. The most prominent
difference is the absence of predicates. 

\defn[Lexicon]
The lexicon consists of:
\begin{enumerate}

    \item \emph{basic symbols}: $\lnot\;\;\land\;\;\lor\;\;\to\;\;\exists\;\;\forall\;\;(\;\;)$
    
    \item \emph{modal operators}: for every accessibility relation $\pi$
    the modal operators $\langle\pi\rangle$ and $[\pi]$

    \item a collection of \emph{constants symbols}. There are two kinds of
    
    constants: constants that denote individual objects and constants that denote
    individual concepts. There is a concept constant symbol $id$, which will be given a special meaning
    in definition \ref{mdcorr}.

    \item a collection of \emph{variable symbols}. Like constants, there are two kinds
    of variables. I'll use lowercase Latin letters $x,y,z$ as object variables and
    lowercase Greek letters $\alpha, \beta, \gamma$ as concept variables.
    
    \item the relation symbols = and $\ne$.
\end{enumerate}

\defn[Term]
A \emph{term} denotes an individual object or concept. The definition of \emph{term}
is as follows:

\begin{enumerate}

    \item Constants and variables are \emph{terms}. A term is an \emph{object term} if it is an
    individual object variable or constant. Similarly for \emph{concept terms}.

    \item If $t$ is a concept term, $\da t$ is an object term. $\da t$ is intended to designate
    the object denoted by $t$, in a particular state.

    \item Nothing else is a term.
\end{enumerate}

\defn[Formula]
A \emph{formula} expresses some fact about the (possibly virtual) reality.
A formula without free variables is called a \emph{sentence}. Sentences are the
things of which we can say that they are true or false.
The definition of formula is as follows:

\begin{enumerate}

    \item  If $t_1$ and $t_2$ are both object terms\footnote{The exclusion of concept terms is intentional.}, 
    then $t_1 = t_2$ and $t_1 \ne t_2$ are atomic formulas.

    \item If $\varphi$ is a formula, then $\lnot \varphi$ is a formula.
    
    \item If $\varphi$ is a formula and $\pi$ an accessibility relation, then $\langle\pi\rangle \varphi$
    and $[\pi] \varphi$ are formulas.
    
    \item If $\varphi$ and $\psi$ are formulas, so are  $(\varphi \land \psi)$,
    $(\varphi \lor \psi)$, $(\varphi \to \psi)$.
    
    \item If $\varphi$ is a formula and $\varrho$ is a variable of either kind, then $\forall \varrho\; \varphi$ and
    $\exists \varrho\; \varphi$ are formulas.
        
    \item If $\varphi$ is a formula, $\varrho$ is a variable of either kind, and $t$ is a term
    of the same kind as $\varrho$, then $\langle \lambda \varrho.\varphi\rangle(t)$ is a formula.
    
    \item Nothing else is a formula.
\end{enumerate}

\subsection{Semantics}

\defn[Augmented Frame]
\label{augmented frame}
The frames we need to build first-order modal models with are enhanced versions of frames used 
for the semantics of propositional modal logic. Let $\Pi$ be a set of accessibility relations.
An \emph{augmented frame} $\F$ is a structure $\langle \G,\{R_\pi | \pi \in \Pi \}, \D_o, \D_c \rangle$
that consists of the following ingredients:

\begin{enumerate}
    \item a non-empty set $\G$ of states. (worlds)

    \item for every $\pi$ in $\Pi$, a binary relation $R_\pi$ on $\G \times \G$.

    \item a non-empty set of objects $\D_o$, called the \emph{domain of the frame}.

    \item $\D_c$ is a non-empty set of functions from $\G$ to $\D_o$, called \emph{individual concepts}.
\end{enumerate}

The domain of an augmented frame is the set of things over which quantifiers can range, no matter
at which state.
$\R$ will be used as shorthand notation for $\{R_r | r \in \Pi \}$.

\defn[Interpretation]
\label{interpretation}
$\I$ is an \emph{interpretation} in an augmented frame
$(\G,\R, \D_o, \D_c )$ if $\I$ is a mapping that assigns:

\begin{enumerate}
    \item to each individual object constant symbol some member of $\D_o$.

    \item to each individual concept constant symbol some member of $\D_c$.
\end{enumerate}

This interpretation gives rise to a \emph{constant domain}, that is, a domain
(of interpreted constants) that is invariable between states.
It is assumed that individual objects and concepts have \emph{unique names}.
In other words, no two different constant symbols denote the same object.
This allows us to use constant symbols to identify objects and concepts and vice versa.
\defn[Model]
\label{model}
A \emph{first-order modal model} is a pair $\M = (\F,\I)$ where
$\F$ is an augmented frame and $\I$ is an interpretation in it. 

\defn[Assignment]
\label{assignment}
Let $\M=(\F,\I)$ be a first-order modal model.
A \emph{assignment} $v$ in the model $\M$ is a mapping that
assigns to each free individual object variable some member of $\D_o$ and
to each free individual concept variable some member of $\D_c$.

\defn[Term evaluation]
\label{term evaluation}
Let $\F = \langle \G,\R, \D_o, \D_c \rangle$ be an augmented frame,
$\M = (\F,\I)$ be a model based on $\F$ and $v$ be an assignment in $\M$.
A mapping $(v * \I)$ is defined, assigning a meaning to each term,
at each possible state. Let $\Gamma \in \G$.

\begin{enumerate}
    \item if $\varrho$ is a variable, $(v * \I)(\varrho,\Gamma) = v(\varrho)$.

    \item if $c$ is a constant symbol, $(v * \I)(c,\Gamma) = \I(c)$.

    \item if $\da t$ is a relativized term, $(v * \I)(\da t,\Gamma) = (v * \I)(t)(\Gamma)$.
\end{enumerate}

To make reading easier, the following special notation is used. Let $\varrho_1, \ldots, \varrho_k$ be variables
of any type, and let $d_1, \ldots, d_k$ be members of $\D_o \union \D_c$, with $d_i \in \D_o$
if the variable $\varrho_i$ is of object type, and $d_i \in \D_c$ if $\varrho_i$ is of concept type.
Then 
\[
\M,\Gamma \Vdash_v \varphi[\varrho_i/d_1, \ldots, \varrho_k/d_k]
\]
abbreviates: $\M,\Gamma \Vdash_{v'} \varphi$ where $v'$ is the assignment that is like $v$
on all variables except $\varrho_1, \ldots \varrho_l$, and $v'(\varrho_1)=d_1, \ldots v'(\varrho_k)=d_k$.

\defn[Truth in a model]
\label{truth definition}
Let $\F = \langle \G,\R, \D_o, \D_c \rangle$ be an augmented frame,
$\M = (\F,\I)$ be a model based on $\F$ and $v$ be a assignment in $\M$.
We now inductively define the notion of a formula $\varphi$ being \emph{satisfied}
(true) in $\M$ at state $\Gamma$ as follows:
\begin{enumerate}   

\item $\M,\Gamma \Vdash_v t_1 = t_2$ iff $(v * \I)(t_1,\Gamma) = (v * \I)(t_2,\Gamma)$.

\item $\M,\Gamma \Vdash_v t_1 \ne t_2$ iff $(v * \I)(t_1,\Gamma) \ne (v * \I)(t_2,\Gamma)$.

\item $\M,\Gamma \Vdash_v \lnot \varphi$ iff
    $\M,\Gamma \not \Vdash_v \varphi$.

\item $\M,\Gamma \Vdash_v (\varphi \land \psi)$ iff
    $\M,\Gamma \Vdash_v \varphi$ and $\M,\Gamma \Vdash_v \psi$.

\item $\M,\Gamma \Vdash_v (\varphi \lor \psi)$ iff
    $\M,\Gamma \Vdash_v \varphi$ or $\M,\Gamma \Vdash_v \psi$.

\item $\M,\Gamma \Vdash_v \varphi \to \psi$ iff
    $\M,\Gamma \not \Vdash_v \varphi$ or $\M,\Gamma \Vdash_v \psi$.

\item $\M,\Gamma \Vdash_v \forall x\;\varphi$ iff
    $\M,\Gamma \Vdash_v \varphi[x/d]$ for all $d \in \D_o$.

\item $\M,\Gamma \Vdash_v \forall \alpha\;\varphi$ iff
    $\M,\Gamma \Vdash_v \varphi[\alpha/d]$ for all $d \in \D_c$.

\item $\M,\Gamma \Vdash_v \exists x\;\varphi$ iff
    $\M,\Gamma \Vdash_v \varphi[x/d]$ for some $d \in \D_o$.

\item $\M,\Gamma \Vdash_v \exists \alpha\;\varphi$ iff
    $\M,\Gamma \Vdash_v \varphi[\alpha/d]$ for some $d \in \D_c$.

\item $\M,\Gamma \Vdash_v [\pi]\varphi$ iff
    for all $\Delta \in \G$, if $\pi(\Gamma,\Delta)$ then $\M,\Delta \Vdash_v \varphi$.

\item $\M,\Gamma \Vdash_v \langle\pi\rangle\varphi$ iff
    for some $\Delta \in \G$, if $\pi(\Gamma,\Delta)$ then $\M,\Delta \Vdash_v \varphi$.

\item $\M,\Gamma \Vdash_v \langle \lambda \varrho.\varphi\rangle(t)$ if 
    $\M,\Gamma \Vdash_v \varphi[\varrho/d]$ where $d=(v*\I)(t,\Gamma)$.
\end{enumerate}

\defn[Modal query]
$\varphi(\varrho_1, \ldots, \varrho_n )$ is a \emph{modal query}, iff

\begin{enumerate}
\item
    $\varphi$ is a wff of the modal language.
\item
    $\varrho_1, \ldots, \varrho_n$ are distinct variables of either kind
\item
    $\varrho_1, \ldots, \varrho_n$ are the only free variables in $\varphi$.
\end{enumerate}

$\varrho_1, \ldots, \varrho_n$ is called the \emph{target list}.

\section{Database and algebra}
We adopt the unnamed conventional perspective of the relational modal, 
which is described in detail in chapter 3 of \cite{alice}.
The unnamed perspective is preferred over the named perspective, because
it's easier to work with in the translation procedure and correspondence proof later in this section.

\subsection{Database}
\defn[Database]

{\bf dom} is a countably infinite set of individual objects.
{\bf relname} is a countably infinite set of relation names.
A \emph{relation scheme} is a relation name (symbol) $R$ along with a positive integer called the \emph{degree} (arity)
of $R$. If $R$ has degree $n$, the $n$ attributes of $R$ are identified by the numbers $1,\ldots,n$.
A \emph{relation instance} $I$, also associated with a degree $n$, is a finite set of $n$-tuples.
\\
\\
\begin{tabular}{llll}
symbol  &   used for\\
\hline
t,u     &   tuple variables\\
a,b,c   &   constant symbols\\
R,S     &   relation names\\
I,J     &   relation instances\\
q       &   queries\\
{\bf R} &   database schema\\ 
{\bf I} &   database instance\\
\end{tabular}

\subsection{Relational Algebra}
Five primitive algebra operators form the \emph{unnamed relational algebra}: projection, selection and cross product, union
and set difference. The sixth operator, intersection, is added because it is the natural algebra counterpart of the
conjunction logical connective.

\begin{defn}[Selection]
\label{selection}
Let $j,k$ be positive integers and $c \in$ {\bf dom}. Then 
$\sigma_{j=c}$ and $\sigma_{j=k}$ are selection operators.
These operators applies to any relation instance $I$ with degree$(I) \ge $ max$\{j,k\}$.
The operator $\sigma_{j=c}$ is defined as follows:
\begin{displaymath}
    \sigma_{j=c}(I) = \{t \in I|t(j)=c\}
\end{displaymath}
producing output of degree$(I)$.
\end{defn}

\defn[Projection]
The projection operator has the form $\pi_{j_1,\ldots,j_n}$ where $j_1, \ldots, j_n$
is a possibly empty sequence of positive integers, possibly with repeats.
This operator takes as input any relation instance with degree $\ge$ max$\{j_1,\ldots,j_n\}$,
and returns an instance with degree $n$, in particular,
\begin{displaymath}
    \pi_{j_1,\ldots,j_n}(I) = \{\langle t(j_1),\ldots,t(j_n)\rangle | t \in I \}
\end{displaymath}.

\defn[Cartesian (cross) product]
Let $I$ and $J$ be two relation instances., with arities $n$ and $m$, respectively.
The \emph{cartesian product} returns a relation instance with a degree of $n+m$ and
is defined as follows
\begin{displaymath}
    I \times J = \{ \langle t(1),\ldots, t(n),u(1),\ldots,u(m)\rangle | t \in I \text{ and } u \in J  \}
\end{displaymath}
The relation instance $\{\langle \rangle\}$ behaves as left and right identity:
\begin{displaymath}
    I \times \{ \langle \rangle \} = \{ \langle \rangle \} \times I = I
\end{displaymath}

Because cross-product is associative, it can be viewed as polyadic operator and written as $I_1 \times \ldots \times I_n$.

\defn[Union-compatible]
Two relations are \emph{union compatible} if they are of the same degree.

\defn[Union]
By adding union to the algebra, it becomes possible to express disjunctive information in algebra expressions.
Let $I$ and $J$ be two relation instances that are union-compatible. The union of $I$ and $J$, noted $I \cup J$,
is defined as follows: 
\begin{displaymath}
    I \cup J = \{ t | t \in I \lor t \in J \}
\end{displaymath}

\defn[Difference]
\label{difference}
Set difference adds negation to the algebra.
Let $I$ and $J$ be two relation instances that are union-compatible. The set difference of $I$ minus $J$, noted $I - J$,
is defined as follows: 
\begin{displaymath}
    I - J = \{ t | t \in I \land t \not \in J \}
\end{displaymath}

\defn[Intersection]
\label{intersection}
Let $I$ and $J$ be two relation instances that are union-compatible. The intersection of $I$ and $J$, noted $I \cap J$,
is defined as follows:
\begin{displaymath}
    I \cap J = \{ t | t \in I \land t \in J \}
\end{displaymath}

\defn[Algebra query]

The \emph{base algebra queries} are inductively defined as follows:
\begin{enumerate}
    \item \emph{Unary singleton constant} : If $c \in$ {\bf dom}, then $\{\langle c \rangle\}$ is a query with $degree$ 1.
    \item \emph{Input relation}: If $R$ is a relation, the expression $R$ is a query with degree equal to $degree(R)$.
\end{enumerate}
The family of \emph{algebra queries} is inductively defined as follows:
\begin{enumerate}
    \item  All base algebra queries are algebra queries.

    \item \emph{Selection}: Let $j,k \le degree(q_1)$ and $c \in$ {\bf dom}.
    If $q_1$ is a algebra query, then $\sigma_{j=c}(q_1)$ and $\sigma_{j=k}(q_1)$ are algebra queries with
    degrees equal to $degree(q_1)$,

    \item \emph{Projection}: If $q_1$ is a algebra query and each $j_1,\ldots,j_n \le degree(q_1)$,
    then $\pi_{j_1,\ldots,j_n}(q_1)$ is a algebra query,
    with degree $n$.

    \item \emph{Cross product}: If $q_1, q_2$ are algebra queries with degrees $n$ respectively $m$,
    then $q_1 \times q_2$ is a algebra query,
    with degree $n+m$.

    \item \emph{Union}: If $q_1, q_2$ are algebra queries that are union compatible; they are of the same degree,
    then $q_1 \cup q_2$ is a algebra query with degree $degree(q_1)$.

    \item \emph{Intersection}: If $q_1, q_2$ are algebra queries that are union compatible,
    then $q_1 \cap q_2$ is a algebra query with degree $degree(q_1)$.

    \item \emph{Difference}: If $q_1, q_2$ are algebra queries that are union compatible,
    then $q_1 - q_2$ is a algebra query with degree $degree(q_1)$.
\end{enumerate}

\section{Translation}
\defn[correspondence]
\label{mdcorr}
Let $\M = \langle \F, \I \rangle$ be a model and $\C$ a bijective mapping that assigns to each concept in $\D_c$ a
number between 1 and $|\D_c|$. Associated with $\M$ are a unique database schema
${\bf R_\M}$ and instance ${\bf I_\M}$ for which the following condition holds:

\begin{enumerate}
    \item ${\bf R_\M} = \{ Sta, Rel, Con, Obj \}$.
    
    \item $Con$ has degree 1 and $Con({\bf I_\M}) = \D_c$.

    \item $Obj$ has degree 1 and $Obj({\bf I_\M}) = \D_o$.

    \item The domain {\bf dom} of the database is $\D_c \cup \D_o \cup \Pi$.
    
    \item The modal language and the database share the same collection of object constant
    symbols.

    \item Unique names are assumed, in particular, the object denoted by a database object constant symbol $c$ is equivalent to it's 
    interpretation in the modal language $\I(c)$.

    \item degree$(Sta) = |\D_c|$

    \item $\Gamma$ is a state in $\G$ iff there is a tuple
    $\langle a_1, \ldots, a_n \rangle$ in $Sta({\bf I_\M})$ such that for every $i \in \{1, \ldots, n\}$
    holds $\I(a_i) = ( v * \I)(\da c)(\Gamma)$, where $\C(c)=i$  and $n =\;$degree$(Sta)$.

    \item there is a concept named $id$ in $\D_c$ such that $Sta : id \to c_1 \ldots c_k$ where $\bigcup_{i=1\ldots k} c_i = \D_c - id$.
    $\C(id)=1$.
    
    \item $Dom(typeCode) = \Pi$. 
    
    \item $Rel$ has degree 3.

    \item for every $\pi \in \Pi$ holds:\\ $\pi(\Gamma,\Delta)$ iff there is a tuple
    $\langle a, b, c \rangle$ in $Rel({\bf I_\M})$ such that 
    $\I(a) = ( v * \I)(\da id)(\Gamma)$ and  
    $\I(b) = ( v * \I)(\da id)(\Delta)$ and  
    $\I(c) = \pi$.
\end{enumerate}

\exmp[]
This example shows a model and it's corresponding database instance:
\\
\\
{\bf Model:}
\\
\\
\xymatrix{
& & *+[F]\txt{id=1\\code=d} \ar@{.>}[lldd]_{COMP}  \ar@{.>}[dd]_{COMP}  \ar@{.>}[rrdd]_{COMP} & & \\
& &                                            & &\\
*+[F]\txt{id=2\\code=a} & & *+[F]\txt{id=3\\code=b} & & *+[F]\txt{id=4\\code=c}\\
}
\\
\\
\\
\\
{\bf Database:}
\\
\\
\begin{tabular}{llll}
Sta & \vline & id & code\\
   \hline
  & \vline  & 1 & d \\
  & \vline  & 2 & a \\
  & \vline  & 3 & b \\
  & \vline  & 4 & c \\
\end{tabular}
\\
\\   
\begin{tabular}{lllll}
Rel & \vline & source & target & typeCode\\
   \hline
  & \vline  & 1 & 2 & COMP \\
  & \vline  & 1 & 3 & COMP \\
  & \vline  & 1 & 4 & COMP \\
\end{tabular}
\\
\\
\begin{tabular}{lll}
Con & \vline & 1\\
   \hline
  & \vline  & id\\
  & \vline  & code\\
\end{tabular}
\\
\\
\begin{tabular}{lll}
Obj & \vline & 1\\
   \hline
  & \vline  & 1\\
  & \vline  & 2\\
  & \vline  & 3\\
  & \vline  & 4\\
  & \vline  & a\\
  & \vline  & b\\
  & \vline  & c\\
  & \vline  & d\\
\end{tabular}

\defn[Formula translation]
\label{ft}
The following translation takes as input a query $\varphi(\varrho_1,\ldots,\varrho_n)$ in the
modal language and results in a relational algebra expression. The translation
consists of a set of syntactic translation rules.
The basic idea is that each atomic subformula, with free variables $\varrho_1,\ldots,\varrho_n$
is translated to a query on $Sta$, $Obj$ and $Con$ that has the following structure:
\\
\\
\begin{tabular}{lllll}
$FT(\varphi(\varrho_1,\ldots,\varrho_n)$ & \vline\; 1              & $\ldots$ & $n$            & id\\
   \hline
                                         & \vline\; $v(\varrho_1)$ & $\ldots$ & $v(\varrho_n)$ & $\Gamma$\\
\end{tabular}
\\
\exmp[no variable query image]
The query image of a translated formula with no variables looks like this:
\\
\\
\begin{tabular}{lllll}
$FT(\da code=b)$ & \vline\;  id\\
   \hline
                 & \vline\;  3
\end{tabular}
\exmp[one variable query image]
The query image of a translated formula with one variables looks like this
\\
\\
\begin{tabular}{lllll}
$FT((\da id=3 \land code=\varrho_1)(\varrho_1))$ & \vline\;  $\varrho_1$ & $id$\\
   \hline
                 & \vline\;  b & 3
\end{tabular}

Conjunctions and disjunctions result in intersections and unions
of queries.
Negation of a query is translated to set difference on $Sta$.
The translation of the existential quantifier is done by translating 
to a query with the quantified variable added to it's target list,
which is later removed by projection of the original target list.
The universal quantifier is translated by translating into the division
of the translation of the remaining subformula of the query, by
the concept domain $Con$ or object domain $Obj$.
The diamond modal operator is translated to a query on $Rel$.
Since the translation doesn't require a specific normal form, 
we can use the dual of the diamond operator to translate the box
operator.
Lambda abstraction is translated using an extra query that captures
the designation of the relativized term.

\begin{enumerate}
\item
Term translations result in attribute index numbers or constants.

\begin{enumerate}
\item
$TT(c) = \text{{\tt '}}c\text{{\tt '}}$, if $c$ an object constant.
\item
$TT(\varrho_k) = k$, if $\varrho_k$ is a variable of either kind.
\item
$TT(\da t) = n + \C(t)$, if $\da t$ is a relativized term. $n$ is
the number of variables of the subformula in the current scope.
\end{enumerate}
\item
Variables result in domain relations.

\begin{enumerate}
\item
$VT(\ro) = \{ \langle \rangle \}$, if $\ro$ is empty.
\item
$VT(\varrho_1, \ldots, \varrho_n) = D_1 \times \ldots \times D_n$, otherwise,
where $D_i$ is the relation $Con$, if $\varrho_i$ is a concept variable,
and $Obj$ if $\varrho_i$ is an object variable, $1 \le i \le n$.
\end{enumerate}

\item
(sub)formula translations are translated to algebra queries.

\begin{enumerate}
\item
$FT(\;(t_1=t_2)(\ro)\;) = \pi_{1,\ldots,n+1}\sigma_{( TT(t_1) = TT(t_2))}(VT(\ro) \times Sta)$
\item
$FT(\;(t_1\ne t_2)(\ro)\;) = \pi_{1,\ldots,n+1}\sigma_{( TT(t_1) \ne TT(t_2))}(VT(\ro) \times Sta)$
\item
$FT(\;\lnot \varphi (\ro)\;) = (VT(\ro) \times \pi_1(Sta) ) - FT( \varphi(\ro) )$
\item
$FT(\;\langle\pi\rangle \varphi (\ro)\;) = \pi_{1,\ldots,n,n+2}\sigma_{(n+4=\pi \land n+1=n+3)}(FT(\varphi(\ro)) \times Rel)$
\item
$FT(\;[\pi] \varphi (\ro)\;) = FT(\lnot \langle\pi\rangle \lnot \varphi (\ro)\;)$
\item
$FT(\;(\varphi \lor \psi)(\ro)\;) = FT( \varphi(\ro) ) \cup FT( \psi(\ro) )$
\item
$FT(\;(\varphi \land \psi)(\ro)\;) = FT( \varphi(\ro) ) \cap FT( \psi(\ro) )$
\item
$FT(\;\exists \varrho\; \varphi(\ro)\;) = \pi_{2,\ldots,n+2}FT( \varrho,\varrho_1, \ldots, \varrho_n )$
\item
$FT(\;\forall \varrho\; \varphi(\ro)\;) = \pi_{2,\ldots,n+2}U - \pi_{2,\ldots,n+2}((VT(\varrho) \times \pi_{2,\ldots,n+2}U)-U)$,
where $U = FT( \varrho,\varrho_1, \ldots, \varrho_n )$
\item
$FT(\langle \lambda \varrho.\varphi \rangle (t)(\ro))=\pi_{2,\ldots,n+2}\sigma_{(1=n+3\land n+2=n+4)}(FT( \varphi(\varrho,\varrho_1, \ldots, \varrho_n)) \times \pi_{TT(t),1}Sta\;)$

\end{enumerate}

\end{enumerate}

\subsection{Examples}

\exmp[Atomic formula, no variables]
Here follows the translation of the variable free query $(\da code = \text{'b'})$.
In this example, the translation of $\times VT(\ro)$ with an empty $\ro$ is given explicitly. In the remaining
examples, I will omit this explicit translation of empty variable lists and directly write $S$ instead of $S \times VT()$.

\begin{displaymath}
\begin{array}[2]{ll}
    FT(\da code = b) & \wrd \pi_{1,\ldots,1}\sigma_{( TT(\da code) = TT(b))}(VT() \times Sta)\\
    & \wrd \pi_{1}\sigma_{2 = \cons{b}}(\{\langle \rangle \} \times Sta)\\
    & \wrd \pi_{1}\sigma_{2 = \cons{b}}Sta\\
\end{array}
\end{displaymath}

\exmp[Diamond operator, no variables]
Here follows the translation of the variable free query $\langle COMP \rangle (\da code = \text{'b'})$.
\begin{displaymath}
\begin{array}[2]{ll}
    FT(\langle COMP \rangle(\da code = b))  & \wrd \pi_{1,\ldots,0,2}\sigma_{(4=\cons{COMP}\land 1=3)}( FT(\da code = b) \times Rel)\\
    & \wrd \pi_{2}\sigma_{(4=\cons{COMP}\land 1=3)}( \pi_{1}\sigma_{2 = \cons{b}}(Sta) \times Rel)\\
\end{array}
\end{displaymath}

\exmp[Box operator, no variables]
\begin{displaymath}
\begin{array}[2]{ll}
    & FT([COMP](\da code = b)) \\
    & \wrd FT(\lnot \langle COMP \rangle \lnot (\da code = b))\\
    & \wrd (VT(\ro) \times \pi_1 Sta) - FT( \langle COMP \rangle \lnot (\da code = b))\\
    & \wrd \pi_1 Sta - (\pi_2\sigma_{(4=\cons{COMP}\land 1=3}( Rel \times FT(\lnot (\da code = b))))\\
    & \wrd \pi_1 Sta - (\pi_2\sigma_{(4=\cons{COMP}\land 1=3)}( Rel \times ((VT(\ro) \times \pi_1 Sta) - FT((\da code = b)(\ro)))))\\
    & \wrd \pi_1 Sta - (\pi_2\sigma_{(4=\cons{COMP}\land 1=3)}( Rel \times (\pi_1 Sta - \pi_{1}\sigma_{2 = \cons{b}}Sta)))\\
\end{array}
\end{displaymath}

\exmp[Predicate abstraction, no variables]
Here follows the translation of the variable free query $\langle \lambda y. \langle COMP \rangle (\da code = y)\rangle (\da code)$.
Note that the first translation step introduces a variable.
\begin{displaymath}
\begin{array}[2]{ll}
    & FT(\langle \lambda y. \langle COMP \rangle (\da code = \varrho)\rangle (\da code)) \\
    & \wrd \pi_{2,\ldots,2}\sigma_{(1=3\land2=4)}(FT(\langle COMP \rangle (\da code = \varrho)(\varrho)) \times \pi_{TT(code)-0,1}Sta)\\
    & \wrd \pi_2\sigma_{(1=3\land2=4)}(\pi_{1,\ldots,1,2}\sigma_{5=\cons{COMP}\land2=5}(FT((\da code = \varrho)(\varrho)) \times Rel) \times \pi_{2,1}Sta)\\
    & \wrd \pi_2\sigma_{(1=3\land2=4)}(\pi_{1,2}\sigma_{5=\cons{COMP}\land2=5}( \pi_{1,2}\sigma_{TT(\da code)=TT(\varrho)}(VT(\varrho) \times Sta) \times Rel) \times \pi_{2,1}Sta)\\
    & \wrd \pi_2\sigma_{(1=3\land2=4)}(\pi_{1,2}\sigma_{5=\cons{COMP}\land2=5}( \pi_{1,2}\sigma_{3=1}(Obj \times Sta) \times Rel) \times \pi_{2,1}Sta)\\
\end{array}
\end{displaymath}

\subsection{Proof of correspondence}

\begin{lem}[]
\label{unionstaisobj}
    Let ${\bf I_\M}$ be a database instance that is associated with a model $\M$.
    Then $\bigcup_{i = 1, \ldots, n} \pi_i Sta({\bf I_\M}) \subseteq Obj({\bf I_\M})$.
\end{lem}
If this was not the case, it would be a violation of definition \ref{mdcorr}, item 4.

\begin{lem}[]
\label{stanotempty}
    Let ${\bf I_\M}$ be a database instance that is associated with a model $\M$.
    Then $Sta({\bf I_\M})$ is not empty.
\end{lem}
This property follows straight from definition \ref{augmented frame}, item 1
and definition \ref{mdcorr}, item 8.

\begin{lem}[]
\label{basecase}
    Fix a model state pair $\M,\Gamma$.
    Then for any two arbitrary terms $t_1, t_2$, assignment $v$ and 
    and object constant $i$, where $i$ identifies the state in $Sta$ such that $\I(i) = (v*\I)(\da id,\Gamma)$,
    the following holds:
    \begin{displaymath}
        (v*\I)(t_1,\Gamma) = (v*\I)(t_2,\Gamma)\text{ iff }\langle v(\varrho_1),\ldots,v(\varrho_n), i \rangle \in FT((t_1=t_2)(\ro))({\bf I_\M})
    \end{displaymath}
    where $\varrho_1, \ldots, \varrho_n$ is the list of variables in $t_1,t_2$.
    \footnote{The object constant symbol $i$ is used on many occasions where $\I(i)$ is more appropriate.
    From the context it should be clear whether the object itself,
    a number or word that identifies a state, or it's (unique) symbol is meant.}
\end{lem}

\begin{proof}
Let $t_1, t_2$ be terms, $v$ an assignment and $u$ the tuple $\langle v(\varrho_1),\ldots,v(\varrho_n), i \rangle$
such that the following holds:
\begin{displaymath}
    u \in FT((t_1=t_2)(\ro))({\bf I_\M})
\end{displaymath}
which by translation step 3(a) of definition \ref{ft} is equal to.
\begin{displaymath}
    u \in \pi_{1,\ldots,n+1}\sigma_{(TT(t_1)=TT(t_2))}(VT(\ro) \times Sta)({\bf I_\M})
\end{displaymath}
Thus the following equivalence is to be proved:
\begin{displaymath}
    (v*\I)(t_1,\Gamma) = (v*\I)(t_2,\Gamma)\text{ iff }u \in \pi_{1,\ldots,n+1}\sigma_{(TT(t_1)=TT(t_2))}(VT(\ro) \times Sta)({\bf I_\M}).
\end{displaymath}

\begin{itemize}
\item
    $t_1$ and $t_2$ are both constants. Since there are no variables, $u$ is the tuple $\langle i \rangle$ and
    $\pi_{1,\ldots,n+1}\sigma_{(TT(t_1)=TT(t_2))} (VT(\ro) \times Sta)({\bf I_\M})$
    is equivalent with
    $\pi_1\sigma_{TT(t_1)=TT(t_2)}Sta({\bf I_\M})$.

    $\Rightarrow$ 
    By definition \ref{term evaluation} item 2, 
    $(v*\I)(t_1,\Gamma) = (v*\I)(t_2,\Gamma)$ iff $\I(t_1) = \I(t_2)$. 
    By the unique names assumption, $t_1$ and $t_2$ are
    the same constant. 
    By definition \ref{mdcorr}, item 5, this constant exists in the database.
    Hence $TT(t_1) = TT(t_2)$ and therefore
    $\sigma_{TT(t_1)=TT(t_2)}Sta({\bf I_\M}) = Sta({\bf I_\M})$.
    By definition \ref{mdcorr} item 9, 
    $u \in \pi_1\sigma_{TT(t_1)=TT(t_2)}Sta({\bf I_\M})$.

    $\Leftarrow$ 
    Suppose that $t_1$ and $t_2$ are different constants.
    Because of the unique names assumption, $\pi_1\sigma_{TT(t_1)=TT(t_2)}Sta({\bf I_\M})$
    is empty. But $u \in \pi_1\sigma_{TT(t_1)=TT(t_2)}Sta({\bf I_\M})$ and 
    $Sta({\bf I_\M})$ is not empty (lemma \ref{stanotempty}).
    Therefore $t_1$ and $t_2$ are the same constant.
    By definition \ref{mdcorr}, item 5, this constant exists in the modal language.
    Hence  $(v*\I)(t_1,\Gamma) = (v*\I)(t_2,\Gamma)$.
\item
    $t_1$ is a constant, $t_2$ is a relativized concept.
    For the sake of readability, let $t_1$ be the constant $a$
    and $t_2$ be the relativized concept $\da c$, with $\C(c)=2$.
    Let $u \in \pi_{1,\ldots,n+1}\sigma_{(TT(t_1)=TT(t_2))} (VT(\ro) \times Sta)({\bf I_\M})$.
    Because there are no variables, this is equivalent with
    $u \in \pi_1\sigma_{\cons{a}=2}(Sta)({\bf I_\M})$,
    which by definition \ref{selection} holds iff $\langle i \rangle \in \{ t[1] | t \in Sta({\bf I_\M}) \land t[2]=a \}$,
    which holds iff there exists a tuple $t$ in $Sta({\bf I_\M})$ such that $t[1]=i$ and $t[2]=a$.
    Because $\I(i) = (v*\I)(\da id,\Gamma)$ and by
    definition \ref{mdcorr} item 8 this holds iff
    $\I(a) = (v*\I)(\da c,\Gamma)$.
    By definition \ref{interpretation}, this holds iff $(v*\I)(a,\Gamma) = (v*\I)(\da c,\Gamma)$.
\item
    $t_1$ is a variable, $t_2$ is a relativized concept.
    For the sake of readability, let $t_1$ be the variable $\varrho$ and $t_2$ be the relativized concept $\da c$, with $\C(c)=k$.
    Let $v$ be any assignment such that 
    $\langle v(\varrho),i \rangle \in FT((\varrho=\da c)(\varrho))$.
    This is translated to $\langle v(\varrho),i \rangle \in \pi_{1,\ldots,n+1}\sigma_{(TT(\varrho) = TT(\da c))}(VT(\varrho) \times Sta)({\bf I_\M})$,
    which is further translated and simplified to the equivalent $\langle v(\varrho),i \rangle \in \pi_{1,2}\sigma_{(1 = k+1)}(Obj \times Sta)({\bf I_\M})$,
    which by definition \ref{selection} is equal to 
    $\langle v(\varrho), i \rangle \in \{t \in \pi_{1,k+1}(Obj \times Sta)({\bf I_\M})|t(1)=t(k+1)\}$.
    Because lemma \ref{unionstaisobj} holds on database instance ${\bf I_\M}$,
    the set $\{t | t \in \pi_{1,k+1}(Obj \times Sta)({\bf I_\M}) \land t(1)=t(k+1)\}$
    is equal to the set $\{ t | t \in \pi_{k,1}Sta ({\bf I_\M}\}$.
    Hence $\langle v(\varrho),i \rangle$ is in $\pi_{k,1}Sta({\bf I_\M})$ iff $v$ assigns to $\varrho$ the object denoted by
    the attribute with index $k$, which corresponds to relativized concept $\da c$.
    By definition \ref{mdcorr} item 8, this holds iff $v(\varrho) = (v*\I)(\da c,\Gamma)$,
    which, in other words, is equal to $(v*\I)(t_1,\Gamma) = (v*\I)(t_2,\Gamma)$.
\item
    the remaining combinations follow from commutativity and transitivity of $=$.
\end{itemize}

\end{proof}

\begin{prop}[Correspondence]
    Fix a model state pair $\M,\Gamma$
    and object constant $i$, such that $\I(i) = (v*\I)(\da id,\Gamma)$
    Then the following holds
    \begin{displaymath}
        \M,\Gamma \Vdash_v \varphi(\varrho_1, \ldots, \varrho_n)\text{ iff } \langle v(\varrho_1),\ldots,v(\varrho_n), i \rangle \in FT(\varphi(\varrho_1, \ldots, \varrho_n))({\bf I_\M})
    \end{displaymath}
\end{prop}

\begin{proof} By induction on the structure of $\varphi$.

\begin{itemize}
\item
    Base case: 
    $\varphi$ is $t_1 = t_2$, where $t_1, t_2$ are object terms.
    This is lemma \ref{basecase}.
\item 
    Case $\lnot \varphi(\ro)$:
    Assume that $\M,\Gamma \Vdash_v \lnot \varphi(\ro)$. This holds\\
    \begin{tabular}{ll}
            iff &$\M,\Gamma \not \Vdash_v \varphi$ (truth definition)\\
            iff &$\langle v(\varrho_1),\ldots,v(\varrho_n), i \rangle \not \in FT(\varphi(\varrho_1, \ldots, \varrho_n))({\bf I_\M})$ (inductive hypothesis)\\
            iff &$\langle v(\varrho_1),\ldots,v(\varrho_n), i \rangle \in (VT(\varrho) \times \pi_1Sta) - FT(\varphi(\varrho_1, \ldots, \varrho_n))({\bf I_\M})$ (def \ref{difference})\\
            iff &$\langle v(\varrho_1),\ldots,v(\varrho_n), i \rangle \in FT(\lnot \varphi(\ro))({\bf I_\M})$ (def \ref{ft} item 3(c)).
    \end{tabular}
\item
    Case $(\varphi \land \psi)(\ro)$:
    Assume that $\M,\Gamma \Vdash_v \varphi \land \psi(\ro)$. This holds \\
    \begin{tabular}{ll}
        iff &$\M,\Gamma \Vdash_v \varphi$ and $\M,\Gamma \Vdash_v \psi$ (truth definition)\\
        iff &$\langle v(\varrho_1),\ldots,v(\varrho_n), i \rangle \in FT(\varphi(\varrho_1, \ldots, \varrho_n))({\bf I_\M})$ (IH)\\
            & and $\langle v(\varrho_1),\ldots,v(\varrho_n), i \rangle \in FT(\psi(\varrho_1, \ldots, \varrho_n))({\bf I_\M})$ (IH)\\
        iff &$\langle v(\varrho_1),\ldots,v(\varrho_n), i \rangle \in FT(\varphi(\varrho_1, \ldots, \varrho_n))({\bf I_\M})$\\
            &$\cap\;FT(\psi(\varrho_1, \ldots, \varrho_n))({\bf I_\M})$ (def \ref{intersection})\\
        iff &$\langle v(\varrho_1),\ldots,v(\varrho_n), i \rangle \in FT((\varphi \land \psi)(\ro))({\bf I_\M})$ (def \ref{ft} item 3(g)).
    \end{tabular}
\item
    Case $(\varphi \lor \psi)(\ro)$: similar to the conjunction case.
\item
    Case $\langle \pi \rangle \varphi(\ro)$:
    Let $id_\Gamma, id_\Delta$ be shorthand notations for $(v*\I)(id,\Gamma)$, $(v*\I)(id,\Delta)$ respectively.
    It is easy to see that the object constant $i$ is equal to $id_\Gamma$.

    $\Rightarrow$ :
    Assume that $\M,\Gamma \Vdash_v \langle \pi \rangle \varphi(\ro)$.
    By the truth definition, there exists a $\Delta$ such that
    $\pi(\Gamma,\Delta)$ and $\M,\Delta \Vdash_v \varphi(\ro)$.
    By the IH, 
    $\langle v(\varrho_1),\ldots,v(\varrho_n), id_\Delta \rangle \in FT(\varphi(\varrho_1, \ldots, \varrho_n))({\bf I_\M})$.
    Since $\pi(\Gamma,\Delta)$ and because of definition \ref{mdcorr} item 12, 
    the tuple $\langle id_\Gamma, id_\Delta, \pi \rangle \in Rel({\bf I_\M})$.
    The crossproduct $FT(\varphi(\varrho_1, \ldots, \varrho_n)) \times Rel$ contains the following attributes:
    $1,\ldots,n$ are $\varrho_1, \ldots, \varrho_n$.
    $n+1$ the id's of the states in which the subformula $\varphi$ is true.
    At index $n+2$ the $Rel$ relation appears in the cross product: $n+2$ holds the source state, $n+3$ the target state and $n+4$ the typeCode.
    Hence $\langle v(\varrho_1),\ldots,v(\varrho_n), id_\Gamma \rangle \in \pi_{1,\ldots,n,n+2}\sigma_{(n+4=\pi \land n+1=n+3)}(FT(\varphi(\varrho_1, \ldots, \varrho_n)) \times Rel)({\bf I_\M})$.
    Since $i$ is equal to $id_\Gamma$, this means that, $\langle v(\varrho_1),\ldots,v(\varrho_n), i \rangle \in FT(\langle \pi \rangle \varphi(\varrho_1, \ldots, \varrho_n))({\bf I_\M})$.
    
    $\Leftarrow$ :
    Assume that $\langle v(\varrho_1),\ldots,v(\varrho_n), i \rangle \in FT(\langle \pi \rangle \varphi(\varrho_1, \ldots, \varrho_n))({\bf I_\M})$.
    Applying translation step 3(d) of definition \ref{ft} gives \\
    $\langle v(\varrho_1),\ldots,v(\varrho_n), i \rangle \in \pi_{1,\ldots,n,n+2}\sigma_{(n+4=\pi \land n+1=n+3)}(FT(\varphi(\varrho_1, \ldots, \varrho_n)) \times Rel)({\bf I_\M})$.
    This means there exist tuples $t,u$ in respectively $FT(\varphi(\varrho_1, \ldots, \varrho_n))({\bf I_\M})$ and $Rel({\bf I_\M})$,
    such that $t[n+1]=u[2]$ and $u[3]=\pi$ and $u[1]=i$. Let $u[2]=id_\Delta$. 
    Since $i=id_\Delta$ and by definition \ref{mdcorr} item 12, $\pi(\Gamma,\Delta)$.
    Since $u[2]=t[n+1]$, also holds
    $\langle v(\varrho_1),\ldots,v(\varrho_n), id_\Delta \rangle \in FT(\varphi(\varrho_1, \ldots, \varrho_n))$.
    Hence, by the IH, $\M,\Delta \Vdash_v \varphi(\ro)$.
    By the truth definition, $\M,\Gamma \Vdash_v \langle \pi \rangle \varphi(\ro)$.
    
\item
    Case $\exists \varrho\; \varphi(\varrho_1, \ldots, \varrho_n)$:
    Let $U$ be the query $FT(\varphi(\varrho, \varrho_1, \ldots, \varrho_n))$.
    The inductive hypothesis states $\M,\Gamma \Vdash_v \varphi(\varrho, \varrho_1, \ldots, \varrho_n)$ iff 
        $\langle v(\varrho), v(\varrho_1),\ldots,v(\varrho_n), i \rangle \in U({\bf I_\M})$.

    $\Rightarrow$ :
    Assume that $\M,\Gamma \Vdash_v \exists \varrho\; \varphi(\varrho_1, \ldots, \varrho_n)$.
   
    By definition \ref{truth definition} of $\exists$,
    $\M,\Gamma \Vdash_v \varphi[\varrho/d](\varrho_1, \ldots, \varrho_n)$ for some $d \in \D_o$\footnote{Replace with $\D_c$ if $\varrho$ is a concept variable.}.
    By the IH, for some $d \in VT(\varrho)$, 
    $\langle d, v(\varrho_1),\ldots,v(\varrho_n), i \rangle \in U({\bf I_\M})$.
    Hence $\langle v(\varrho_1),\ldots,v(\varrho_n), i \rangle \in \pi_{2,\ldots,n+2}FT(\varphi(\varrho, \varrho_1, \ldots, \varrho_n))({\bf I_\M})$.
    
    $\Leftarrow$ :
    Assume that $\langle v(\varrho_1),\ldots,v(\varrho_n), i \rangle \in \pi_{2,\ldots,n+2}FT(\varphi(\varrho, \varrho_1, \ldots, \varrho_n))({\bf I_\M})$.
    For the sake of contradition, suppose that 
    $\M,\Gamma \Vdash_v \not \exists \varrho\; \varphi(\varrho_1, \ldots, \varrho_n)$.
    By definition \ref{truth definition},
    $\M,\Gamma \not \Vdash_v \varphi[\varrho/d](\varrho_1, \ldots, \varrho_n)$ for some $d \in \D_o$.
    By the IH and dual, for no $d \in \D_o$, 
    $\langle d, v(\varrho_1),\ldots,v(\varrho_n), i \rangle \in U({\bf I_\M})$.
    Hence for all $d \in \D_o$, 
    $\langle d, v(\varrho_1),\ldots,v(\varrho_n), i \rangle \not \in U({\bf I_\M})$
    and hence $\langle v(\varrho_1),\ldots,v(\varrho_n), i \rangle$ $\not \in$
    $\pi_{2,\ldots,n+2}FT(\varphi(\varrho, \varrho_1, \ldots, \varrho_n))({\bf I_\M})$.
    But this is a contradiction, so 
    $\M,\Gamma \Vdash_v \exists \varrho\; \varphi(\varrho_1, \ldots, \varrho_n)$.
    
\item
    Case $\forall \varrho\; \varphi(\varrho_1, \ldots, \varrho_n)$:
    Let $U$ be the query $FT(\varphi(\varrho, \varrho_1, \ldots, \varrho_n))$ and
    let $A$ be the query $VT(\varrho) \times \pi_{2,\ldots,n+2}U$.
    The inductive hypothesis states $\M,\Gamma \Vdash_v \varphi(\varrho, \varrho_1, \ldots, \varrho_n)$ iff 
        $\langle v(\varrho), v(\varrho_1),\ldots,v(\varrho_n), i \rangle \in U({\bf I_\M})$.

    $\Rightarrow$ :
    Assume that $\M,\Gamma \Vdash_v \forall \varrho\; \varphi(\varrho_1, \ldots, \varrho_n)$.
    
    By definition \ref{truth definition} of $\forall$, $\M,\Gamma \Vdash_v \varphi[\varrho/d](\varrho_1, \ldots, \varrho_n)$ for all $d \in \D_o$\footnote{Replace with $\D_c$ if $\varrho$ is a concept variable.}.
    Hence, by the IH, for all $d \in \D_o$, 
    $\langle d, v(\varrho_1),\ldots,v(\varrho_n), i \rangle \in U({\bf I_\M})$ (*).
    It is now easy to see that $A({\bf I_\M}) = U({\bf I_\M})$. Hence $(A-U)({\bf I_\M}) = \emptyset$.
    Hence $\pi_{2,\ldots,n+2}U - \pi_{2,\ldots,n+2}(A)-U)({\bf I_\M}) = \pi_{2,\ldots,n+2}U({\bf I_\M})$.
    Since (*), 
    $\langle v(\varrho_1),\ldots,v(\varrho_n), i \rangle \in \pi_{2,\ldots,n+2}U - \pi_{2,\ldots,n+2}((VT(\varrho) \times \pi_{2,\ldots,n+2}U)-U)({\bf I_\M})$
    By definition \ref{ft}, 
    $\langle v(\varrho_1),\ldots,v(\varrho_n), i \rangle \in FT(\forall \varrho\; \varphi(\varrho_1, \ldots, \varrho_n))({\bf I_\M})$
    
    $\Leftarrow$ :
    Assume that $\langle v(\varrho_1),\ldots,v(\varrho_n), i \rangle \in FT(\forall \varrho\; \varphi(\varrho_1, \ldots, \varrho_n))({\bf I_\M})$.
    By definition \ref{ft}, 
    $\langle v(\varrho_1),\ldots,v(\varrho_n), i \rangle \in \pi_{2,\ldots,n+2}U - \pi_{2,\ldots,n+2}((VT(\varrho) \times \pi_{2,\ldots,n+2}U)-U)({\bf I_\M})$.
    For the sake of contradiction, suppose that $\M,\Gamma \not \Vdash_v \forall \varrho\; \varphi(\varrho_1, \ldots, \varrho_n)$.
    By definition \ref{truth definition}, there exists a $d \in \D_o$ such that
    $v(\varrho) = d$ and $\M,\Gamma \not \Vdash_v \varphi(\varrho, \varrho_1, \ldots, \varrho_n)$.
    By the IH, $\langle v(\varrho), v(\varrho_1),\ldots,v(\varrho_n), i \rangle \not \in U({\bf I_\M})$.
    Hence $\langle v(\varrho), v(\varrho_1),\ldots,v(\varrho_n), i \rangle \in (A - U)({\bf I_\M})$
    and hence $\langle v(\varrho_1),\ldots,v(\varrho_n), i \rangle \in \pi_{2,\ldots,n+2}(A - U)({\bf I_\M})$.
    and hence 
    $\langle v(\varrho_1),\ldots,v(\varrho_n), i \rangle \not \in \pi_{2,\ldots,n+2}U - \pi_{2,\ldots,n+2}((VT(\varrho) \times \pi_{2,\ldots,n+2}U)-U)({\bf I_\M})$
    which is a contradiction.
    Therefore $\M,\Gamma \Vdash_v \forall \varrho\; \varphi(\varrho_1, \ldots, \varrho_n)$.
    
\item
    Case $\langle \lambda \varrho.\varphi \rangle (t)(\varrho_1, \ldots, \varrho_n)$:

    Assume that $\M,\Gamma \Vdash_v \langle \lambda \varrho.\varphi \rangle (t)(\varrho_1, \ldots, \varrho_n)$.

    By the truth definition, this holds iff
    $\M,\Gamma \Vdash_v \varphi[\varrho/d]$ where $d = (v*\I)(t,\Gamma)$.
    In other words,
    $\M,\Gamma \Vdash_{v'} \varphi$ where $v'$ is $v$ except $v'(\varrho) = (v*\I)(t,\Gamma)$ (*).

    By the IH and because $\M,\Gamma \Vdash_{v'} \varphi$, this holds iff
    $\langle v'(\varrho_1),\ldots,v'(\varrho_n) \rangle \in FT( \varphi(\varrho_1, \ldots, \varrho_n))({\bf I_\M})$ ($*^2$).
    Since in subformula $\varphi$, $\varrho$ is an unbound variable, by definition \ref{ft} item 2b,
    $FT( \varphi(\varrho, \varrho_1, \ldots, \varrho_n)) = VT(\varrho) \times FT( \varphi(\varrho_1, \ldots, \varrho_n))$.
    Thus ($*^2$) holds iff
    $\langle v'(\varrho), v'(\varrho_1),\ldots,v'(\varrho_n) \rangle \in FT( \varphi(\varrho,\varrho_1, \ldots, \varrho_n))({\bf I_\M})$
    for any arbitrary valuation $v'$ of $\varrho$.
    Since (*), this holds iff 
    $\langle v'(\varrho), v'(\varrho_1),\ldots,v'(\varrho_n) \in$
    $\pi_{1,\ldots,n+2}\sigma_{(1=n+3\land n+2=n+4)}$
    $(FT( \varphi(\varrho,\varrho_1, \ldots, \varrho_n)) \times \pi_{TT(t),1}Sta)({\bf I_\M})$ ($*^3$).
    The last step explained: $n+2 = n+4$ is a join condition on the state identifiers: select only tuples with matching states.
    The condition $1=n+3$ ensures that $v'(\varrho)$ is equal to attribute with index $TT(t)$, which by definition \ref{ft} item 1c,
    is $\C(t)$.
    In other words, only records are selected where $v'(\varrho)$ is equal to the value of attribute $\C(t)$ in $Sta$.
    By definition \ref{mdcorr} item 8, this means that if and only if
    $v'(\varrho) = (v'*\I)(t,\Gamma)$, the tuple is present in the query image.
    
    Finally, because $v'$ is $v$ except $v'(\varrho) = (v*\I)(t,\Gamma)$ and by definition \ref{ft}, ($*^3$) holds iff
    $\langle v(\varrho_1),\ldots,v(\varrho_n) \rangle \in FT( \langle \lambda \varrho.\varphi \rangle (t)(\varrho_1, \ldots, \varrho_n))({\bf I_\M})$.    
    
\end{itemize}

\end{proof}

\bibliographystyle{alpha}
\bibliography{bibfile}

\end{document}